# Design considerations for future DAΦNE upgrades

D. Alesini, G. Benedetti, M.E. Biagini, C. Biscari, R. Boni, M. Boscolo, A. Clozza, G. Delle Monache, G. Di Pirro, A. Drago, A. Gallo, A. Ghigo, S. Guiducci, M. Incurvati, E. Levichev, C. Ligi, F. Marcellini, G. Mazzitelli, C. Milardi, L. Pellegrino, M.A. Preger, P. Raimondi, R. Ricci, U. Rotundo, C. Sanelli, M. Serio, F. Sgamma, B. Spataro, P. Piminov, A. Stecchi, A. Stella, F. Tazzioli, C. Vaccarezza, M. Vescovi, M. Zobov

Presented by A. Gallo

ABSTRACT

The Frascati Φ-Factory DAΦNE has been delivering luminosity to the KLOE, DEAR and FINUDA experiments since year 2000. Since April 2004 the KLOE run has been resumed and recently peak luminosity of *$1.0 \cdot 10^{32} cm^{-2} s^{-1}$* and integrated luminosity of *6.2 $pb^{-1}$/day* have been achieved. The scientific program of the three high-energy experiments sharing DAΦNE operation will be completed approximately by the end of year 2006. A scientific program for DAΦNE beyond that date has not been defined yet and it is matter of discussion in the high-energy physics and accelerator physics communities.

In this paper we present some future scenarios for DAΦNE, discussing the expected ultimate performances of the machine as it is now and addressing the design for an energy and/or luminosity upgrade. The options presented in the following are not exhaustive and they are intended to give a glance of what is doable using the existing infrastructures.

## 1. Expected ultimate performances of DAΦNE with the present hardware

The histories of the achieved peak and daily integrated luminosities at DAΦNE since, respectively, years 2000 and 2002 are shown in Fig. 1. Even though the progress over the years has been continuous and substantial, we believe that a significant further improvement in terms of peak and integrated luminosities is still possible.

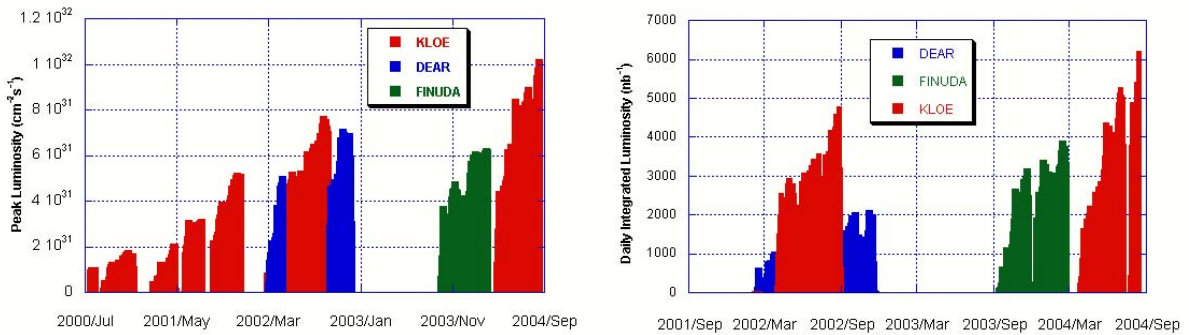

Fig. 1: History of peak and daily integrated luminosities at DAΦNE

The improvement expectations mainly rely on:

- Implementing of a lattice providing negative momentum compaction factor $\alpha_c$ to shorten the bunch and decrease the vertical beta-function $\beta_y^*$ at the Interaction Point (IP);
- Moving the betatron tunes $\nu_{x,y}$ towards the integer to reduce the beam-beam induced blow-up of the bunches;
- Increasing the beam currents by improving the beam dynamics and the performances of the active feedback systems.

Decreasing the $\beta_y^*$, i.e. the value of the vertical beta-function at the IP, is beneficial to the luminosity because it results in a reduction of the vertical size of the bunch $\sigma_y^*$ and of the linear tune shift parameter $\xi_y$ which is an indicator of the strength of the beam-beam effect. However, the beta-function has a parabolic shape around the IP, and the parabola is such that the length of the region where the beta-function remains small is comparable to the $\beta_y^*$ value itself. Accordingly, the $\beta_y^*$ value can not be reduced much below the bunch length value $\sigma_z$ to avoid a geometrical reduction of the luminosity known as "hourglass effect" [1].

The measured bunch length $\sigma_z$ as a function of the bunch current $I_b$ for both DAΦNE rings is shown in Fig. 2. The bunch lengthens with the current because its interaction with the surrounding vacuum chamber generates a self-induced e.m. field which is opposite in phase with respect to the longitudinal focusing RF field provided by the RF system. The bunch lengthens more in the electron ring because of the presence of some extra discontinuities (the ion clearing electrodes) in the vacuum chamber. Presently DAΦNE is operating with $\beta_y^* \approx 1.9\ cm$ and $I_b \approx 10\ mA$, at the threshold of the hourglass effect.

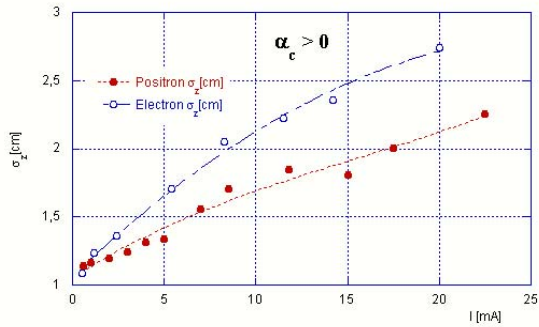 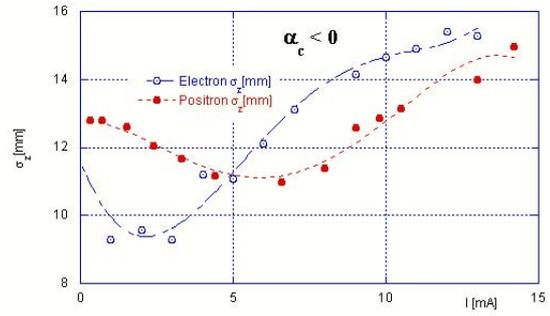

Fig. 2: Bunch length for positive $\alpha_c$    Fig. 3: Bunch length for negative $\alpha_c$

A very effective way to shorten the bunch is to implement a lattice with negative momentum compaction factor $\alpha_c$ [2]. The momentum compaction is the ratio between the relative closed orbit elongation and the relative energy deviation of a particle in a storage ring. In a standard ring ($\alpha_c > 0$) the most energetic particles travel a longer closed orbit. However, if $\alpha_c < 0$ the bunch self induced field is typically in-phase with the external RF field, and the bunch tends to shorten with increasing current. Due to other effects, such as the increase of the bunch energy spread with the current, after reaching a minimum the bunch starts increasing again. Negative momentum compaction lattices have been tested in DAΦNE during machine study shifts. The measured bunch length as a function of the current for the two rings is reported in Fig. 3. Again, the different behavior of the two rings can be explained in terms of different wake-fields, but the bunch lengths remain below 1.5 cm up to bunch current of 15 mA. The potentiality of this kind of lattice is evident, but retuning the machine and all the feedback systems to optimize the luminosity in this configuration will require many machine development shifts. The expected luminosity gain after a complete machine retuning is in the 25÷50 % range.

The beam-beam effect, which is especially severe in the low-energy colliders, is also limiting the luminosity performances of DAΦNE. The primary effect induced by the beam-beam interaction is the blow-up of the transverse (in particular the vertical) size of the bunches of one beam as function of the current in the bunches of the other one.

From beam-beam simulations as well as from experimental data from other colliders we know that a way to limit the beam-beam induced blow-up is to work with betatron tunes close to the integer. According to numerical simulation of a tune scan, this is true also in the

DAΦNE case [3]. Presently we are running the machine with tune fractional parts around $v_x \approx 0.10$, $v_y \approx 0.18$ with a small asymmetry between the two beams. Since any storage ring is unstable at integer betatron tunes, working close to the integer is critical and requires a very fine tune-up of the machine linear and non linear corrections.

The reduction of the DAΦNE betatron tunes towards lower values has already started as a slow, adiabatic process requiring a machine re-optimization at every new small step. We believe that pushing this process further in this direction is worthwhile.

In the present DAΦNE operation the typical multibunch currents in collision after injection are $I^+ \approx 1.0\ A$ and $I^- \approx 1.2\ A$ in $\approx 100$ bunches. Presently the main limitation is in the positron current and it is due to a horizontal multibunch instability causing saturation in the injection, high beam-induced background in the detector and spoiling the uniformity of the bunch train. The origin of this instability is not well understood, but the threshold is slowly increasing with time because of continuous improvements in the setting-up of the bunch-to-bunch feedback systems and non linear correction adjustments. Machine study shifts dedicated to the beam dynamics are needed to better understand and cure these effects in order to increase the current in collision. The progress in this field has been continuous and the colliding currents achieved, in spite of the intrinsic sensitivity of the beam dynamics at low energy, are already comparable with those obtained at the B-factories.

If all the tasks indicated in this paragraph will be pursued during the next two years of operation, the goal of putting in collision multibunch currents in the *1.5÷2.0 A* range to double the present peak and daily integrated luminosities seems realistic. We believe that these numbers represent the DAΦNE potentiality with the present hardware.

## 2. Minimal changes for Energy upgrade from the Φ resonance to the n-nbar threshold

The minimal DAΦNE upgrade to operate the machine at energies from the Φ resonance (0.51 GeV/beam) to the threshold of the n-nbar production (1.1 GeV/beam) requires essentially new dipole magnets fitting the existing vacuum chamber and providing up to 2.4 T magnetic field in the gap [4]. In this way the layout of the machine is preserved. Furthermore, new superconducting quadrupoles housed in the experimental detector to be powered for variable beam energies have to be designed for the low beta insertion. The other existing quadrupoles and their power supplies are basically compatible with 1.1 GeV operation, while only an optimization of the lattice to prevent their saturation is needed. The other existing machine subsystems (such as the vacuum system, the RF, the bunch-to-bunch feedbacks, …) are basically compatible with this option.

A 2D model of a C-shaped dipole for the DAΦNE energy upgrade has been designed. In this special design the magnetic pole tips are made of a special high saturation iron alloy named Hyperco® to reach the required B-field in the gap. The obtained preliminary results show that in principle the required dipoles are feasible, but more work is needed to get a reliable design providing the required field and field quality at any energy in the specified range.

The main machine parameters at the Φ and n-nbar threshold energies are reported in Table 1, columns 2 and 3. Since the luminosity naturally increases with the energy, a peak value $L_{pk} = 1 \cdot 10^{32}\ cm^{-2}s^{-1}$ at the energy of the n-nbar threshold can be obtained with only $\approx 0.5\ A$ of total current in 30 bunches, and with a Touschek lifetime larger than 4 hours. No significant differences are expected for the operation at the Φ energy since the hardware and the machine layout basically remain the same.

Concerning injection, there are two main options: upgrade the DAΦNE linac for full energy injection (without damping ring) [5] or preserve the present injection system (including the damping ring) implementing an energy ramping scheme in the main rings [6]. The energy ramping option requires a synchronized control of the magnet power supplies that is allowed by the existing hardware. This option does not allow topping-up injection in the high energy operation. On the other side, the linac upgrade option surely allows a faster and more flexible injection procedure, but it is far more expensive and requires the upgrade also of the kickers and septum magnets in the ring.

## 3. A new flexible collider for both Energy (up to the n-nbar threshold and beyond) <u>and</u> Luminosity (up to $10^{33}$ cm$^{-2}$s$^{-1}$ at the Φ resonance) upgrade

If a significant increase of the luminosity at the Φ energy is required together with the capability of running at higher energies, the collider has to be completely redesigned and rebuilt. The basic guidelines of a design matching these requirements are drawn in this paragraph.

Table 1:
upgraded DAΦNE and flexible collider parameters at Φ and n-nbar threshold energies

|  | Minimal DAΦNE upgrade | | New flexible collider | |
| --- | --- | --- | --- | --- |
| Energy [GeV] | 0.51 | 1.1 | 0.51 | 1.1 |
| B-field central pole [T] | 1.1 | 2.4 | 2.67 | 2.92 |
| B-field lateral poles [T] | --- | --- | -1.48 | 1.64 |
| Total Current [A] | 1 - 2 | 0.5 | 3 | 0.5 |
| Luminosity [$10^{32}$cm$^{-2}$s$^{-1}$] | 2 | 1 | 10 | 1 |
| N bunches | 100 | 30 | 100 | 30 |
| Current/bunch [mA] | 10-20 | 17 | >20 | 17 |
| Synchrotron integral $I_2$ [m$^{-1}$] | 9.7 | 5.9 | 17.5 | 4.5 |
| Radiation damp. rate [s$^{-1}$] | 25 | 160 | 45 | 115 |
| Energy loss/turn Uo [keV] | 9.3 | 125 | 17 | 110 |

Any upgrade design of DAΦNE as a Φ-factory has to start from an increase of the machine radiation damping rate. In fact, the physics of the beam-beam effect, extensively investigated both theoretically and experimentally, shows that fast radiation damping rates are essential to limit the beam-beam induced vertical blow-up and increase the achievable luminosity. The qualitative explanation of this result is quite intuitive: the faster the damping rate, the shorter the time needed by a particle to loose the "memory" of any experienced perturbation including those coming from beam-beam interaction.

In a storage ring the horizontal damping rate $\alpha_x$ and the energy loss per turn $U_0$ grow respectively with the 3$^{rd}$ and the 4$^{th}$ power of the beam energy $E$ through the synchrotron integral $I_2$ defined as the integral of $1/\rho^2(s)$ over the ring, where ρ(s) is the local bending radius. Increasing $I_2$ is particularly important at low energies, where the damping rate is smaller. A possible way to do that is to divide a bending magnet in 3 pieces, as shown in Fig. 4. The solution is such that a total 30° bending angle is obtained at both 0.5 GeV and 1.1 GeV energies by inverting the polarity in the 2 lateral parts of the magnet and retuning the B-field by a small amount. As shown in Table 1, where a summary of the machine parameters at 0.51 GeV and 1.10 GeV is presented in columns 4 and 5, $I_2$ is ≈ 4 times larger at low energy giving a damping rate only a factor 2.5 smaller with respect to the high energy case. The damping rate at low energy is also almost doubled with respect to the DAΦNE present value, which is very promising for the luminosity performances.

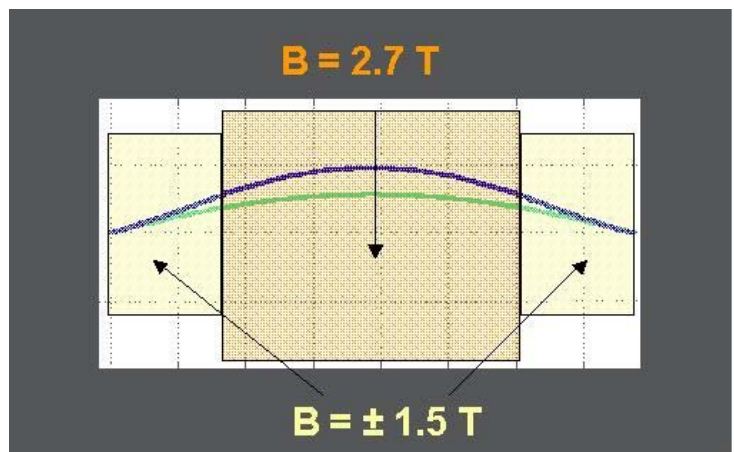
Fig. 4: 3-pieces dipole cell

A total of 12 bending magnets, each made of 3 pieces, are needed for each ring. Due to the high B-field values assumed, the dipoles must be of superconducting type. Energies higher than 1.1 GeV are also accessible (up to the J/ψ or even the τ) provided the B-field in the dipoles and in all other magnets can be increased proportionally.

## 4. A super Φ-factory for Luminosities exceeding $10^{34}$ cm$^{-2}$s$^{-1}$

In this paragraph we summarize the study of a new Φ-factory fitting the existing DAΦNE buildings and pushing the design luminosity at the limit of the accelerator physics state of the art [7]. The ultra-high luminosity design is based on a mix of standard and new concepts, the most important ones being:

- Strong radiation emission to increase radiation damping;
- Large and negative momentum compaction lattice;
- Strong RF Focusing scheme to get bunch length in the mm scale.

The importance of enhancing the radiation emission and the potentiality of the negative momentum compaction factor have been already illustrated in the previous paragraphs. The basic "wiggling" cell shown in Fig. 5 made of a sequence of inward and outward bending dipoles provides both large radiation damping and negative momentum compaction. Due to partial compensation of positive and negative dipoles, the total bending angle of one cell is small, and a large number of cells (i.e. a large number of dipoles) can be used to close the machine.

The momentum compaction $\alpha_c$ is given by the integral of the dispersion function $D(s)$ divided by the local bending radius $\rho(s)$. Being the signs of $D(s)$ and $\rho(s)$ opposite in the cell, $\alpha_c$ is naturally negative and large in this structure. A large $\alpha_c$ is necessary to keep the bunch short by implementing the strong RF focusing scheme.

To make a substantial step in the luminosity is necessary to decrease by about one order of magnitude the vertical beta-function at the IP $\beta_y^*$ passing from cm to the mm scale. To do this, as discussed in paragraph 1, the bunch length must be reduced to about the same value to avoid the hourglass effect. Recently, a novel technique called Strong RF Focusing (SRFF) has been proposed to meet this requirement [8]. By combining a very large RF gradient with a large momentum compaction factor, the bunch length can be modulated along the ring. The bunch length has its maximum in the RF section, and the lattice can be tuned in such a way that the bunch length is minimum at the IP. This condition requires that the two portions of the ring delimited by the RF and the IP contribute equally to the total momentum compaction of the ring.

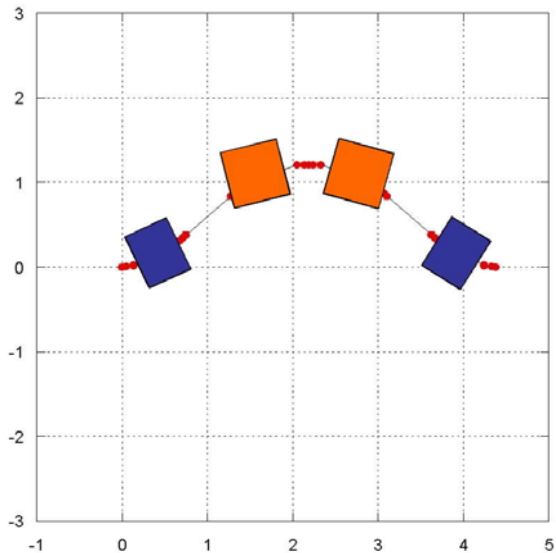
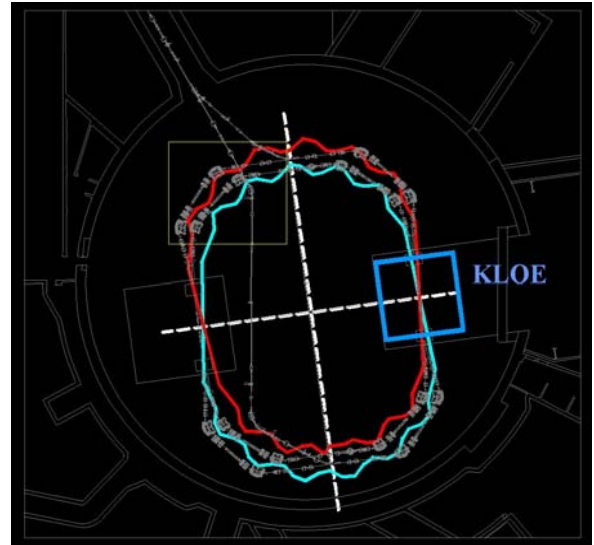

Fig. 5: Wiggling cell      Fig. 6: Layout of the super Φ-factory

It may be seen that even with large $\alpha_c$ values (of the order of $10^{-1}$), the voltage needed to produce sizeable variations of the bunch length along the ring are of the order of 10 MV, a very large value for a 100 m long ring which surely requires a very efficient superconducting RF system.

The main advantage of the SRFF scheme is that the bunch is not short everywhere in the ring. This gives the possibility of placing the impedance generating elements (such as injection and correction kickers, bellows, beam position monitors, …) as much as possible close to the RF section where the bunch is longest. The amplitude of the generated wakefields can be minimized and the bunch can be kept short at the IP up to the nominal operating current (of the order of 15 mA/bunch). Numerical simulations based on the short range wake model of DAΦNE show that this result is achievable.

Table 2: super Φ-factory parameters

| | |
|---|---|
| Total length L [m] | 105 |
| Energy [MeV] | 510 |
| RF frequency $f_{RF}$ [MHz] | 497 |
| RF voltage $V_{RF}$ [MV] | 10 |
| Horiz. emittance $\varepsilon_x$ [μ rad] | 0.26 |
| Vert. emittance $\varepsilon_y$ [μ rad] | 0.002 |
| Momentum compaction $\alpha_c$ | - 0.165 |
| Horiz. beta @ IP $\beta_x^*$ [m] | 0.5 |
| Vert. beta @ IP $\beta_y^*$ [mm] | 2.0 |
| N of particle / bunch | $5\ 10^{10}$ |
| Harmonic number h | 180 |
| Lum./bunch [$cm^{-2}\ sec^{-1}$] | $9\ 10^{31}$ |
| Lum. Tot. [$cm^{-2}\ sec^{-1}$] | $\sim 10^{34}$ |

The new machine layout superimposed to the present one in the DAΦNE hall is shown in Fig. 6, while the main parameters are summarized in Table 2.

The SRFF principle, which is essential to reach the highest luminosities, has never been demonstrated and studied experimentally. Many aspects of beam physics (such as Touschek lifetime, dynamic aperture, beam-beam, coherent synchrotron radiation emission, …) need to be investigated in more detail to establish whether or not a collider may efficiently work in this regime. In order to add reliability to a design based on this scheme, an SRFF experiment to be carried out at DAΦNE has been proposed [9]. A high momentum compaction lattice for DAΦNE has been designed, while an extra SC RF cavity to be temporarily installed in the FINUDA interaction region is under design. Bunch lengths varying from 1.5 to 3 mm along the ring will be obtained. To reduce size and cost, the cavity design has been based on the existing 1.3 GHz, 9-cells TESLA cavities. According to this proposal, the experiment will be completed by the end of 2006 and will give the first SRFF experimental observation, together with other useful experimental results on the impact of this regime on the beam dynamics and on the bunch-by-bunch feedback systems. The experiment cost estimate is ≈1 M€ mainly for the construction of the new SC cavity and cryostat, and to the upgrade of the DAΦNE cryoplant to produce 2 K liquid Helium. The experiment has not been funded yet.

## 5. Conclusions

DAΦNE is running regularly for the KLOE, FINUDA and DEAR/SIDDARTHA experiments, with a continuous improvement of its performances and reliability. The scientific program of the experiments should be completed by the end of 2006 and a new high-energy scientific program beyond that date has not been yet defined. Different upgrade options of the collider fitting the existing infrastructures have been presented: energy upgrade to reach the n-nbar threshold with minimal changes, energy and luminosity upgrade with a new flexible machine, a new super Φ-factory to increase the luminosity by 2 orders of magnitude.
The LNF high-energy physics and the accelerator communities are working together to refine these proposals and converge to a new common enterprise renewing the well-established tradition of these laboratories.